# Molecular Dynamics Simulations of the $O_2^-$ Ion Mobility in Dense Neon Gas

A.F. Borghesani
CNISM unit
Department of Physics & Astronomy
University of Padua, Italy
armandofrancesco.borghesani@unipd.it

F. Aitken
University Grenoble Alpes, CNRS, G2ELab
Grenoble, France
frederic.aitken@ g2elab.grenoble-inp.fr

*Abstract*— We report here the results of Molecular Dynamics simulations of the drift mobility of negative oxygen ions in very dense neon gas in the supercritical phase. The simulations relatively well reproduce the trend of the experimental data. The rationalization of the mobility behavior as a function of the gas density is given in terms of the number of atoms correlated in the first solvation shell around the ion.

## I. Introduction

Electron attachment to electronegative molecules may lead to the formation, under suitable conditions, of negative molecular ions that play an important role in the physics and chemistry of plasmas. The study of these ions is important, among others, for industrial applications (e.g., plasma etching of semiconductors) and for the environment science (e.g., physics and chemistry of the ionosphere). There is also a long standing interest on ionic transport in gases because of the possibility of gathering knowledge on the fundamental interactions of the charge carriers with the gas atoms.

The focus of the experiments on the transport properties of negative ions in gases is typically restricted to very low density non-polar gases [1] in order to determine the ion-atom/molecule interaction potentials. Very few studies are devoted to ionic transport in dense gases which should provide the bridge linking the kinetic- to the hydrodynamic regime that rules the charge transport in the dielectric liquids. Moreover, negative ions are difficult to produce because electron attachment leads to autodissociation of the molecule at very low density. However, at high density, collisions with the atoms of the gas may stabilize the ions that can be drifted under the action of an electric field.

Except for pioneeristic $O_2^-$ mobility studies in dense He gas [2,3], we have systematically investigated the drift mobility of $O_2^-$ in several noble gases [4-6], thereby raising more questions than we solved. Actually, we have shown that neither the kinetic theory in an intermediate density range nor the Stokes law with a constant hydrodynamic radius in the hydrodynamic regime at high density are able to reproduce the experimental data in supercritical nonpolar gases. Only close to the critical point of Ar [6] and Ne [7] the behavior of the mobility as a function of the gas density has been explained within the hydrodynamic-regime model by including the electrostriction phenomenon that leads to the buildup of a *solventberg* around the ions. Moreover, in Ar the contribution of the long-range correlations close to the critical point is to be taken into account [6]. The situation is better for positive He ions [8].

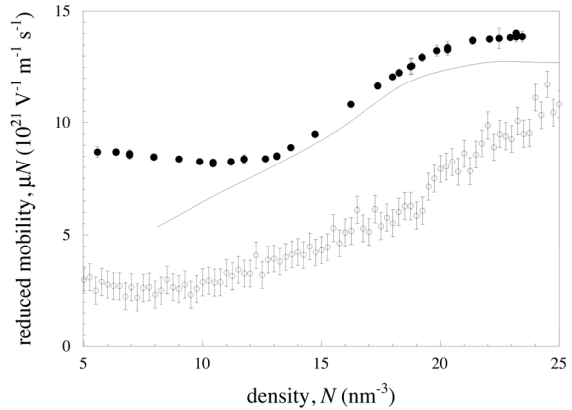

Fig. 1. Density-normalized mobility of $O_2^-$ in dense Ne gas at T=45 K [4] (solid points). MD (open points). Solid line: electrostriction model [7].

The experimental results of the density-normalized mobility $\mu N$ for T=45 K have already been published [4] and are shown in Fig.1 as a function of the gas density $N$. $\mu N$ at (relatively) low density is approximately constant and shows a small dip just below the critical density $N_c$= 14.4 nm$^{-3}$ (the critical temperature is $T_c$=44.4 K). Finally, it again increases with a further density increase. At present, an overall description of the behavior is not at hand for several reasons. First of all, in the low density region there is the lack of the ion-atom interaction potential from which a scattering cross-section can be derived to calculate the mobility according to the Kinetic Theory [10].

On the other hand, hydrodynamics can quite well reproduce the experimental data only if a density dependent hydrodynamic radius is introduced in the Stokes formula. It has been suggested [7] that an empty bubble surrounded by a local density increase is formed as a consequence of the competitive balance between short-range repulsive exchange- and long-range polarization forces between ion and host gas atoms. The cavity radius is assumed to be the ion hydrodynamic radius and the Stokes formula is used to compute the mobility. This approach has given a reasonable agreement with the experimental data, as shown by the solid line in Fig. 1. Nonetheless, the low-density range is not reproduced because it is outside the validity range of the hydrodynamic approach.

Owing to the lack of a full-fledged theory of ion transport, we have been carrying out some Molecular Dynamics (MD) simulations for $O_2^-$ in Ar gas [9] with the hope of elucidate the mechanisms responsible for the observed behavior. We report here the preliminary results obtained for $O_2^-$ ions in dense, supercritical Ne gas at $T$=45 K, shown in Fig.1. In spite of the limitations of MD, the experimental values are satisfactorily reproduced within a factor 2 and the general trend with density is relatively well described. A detailed anaysis of the simulations will be presented next.

## II. MD DETAILS

We have implemented an equilibrium MD method that gives the ion diffusion coefficient $D$ directly and the ion mobility $\mu$ indirectly by using the Nernst-Townsend-Einstein equation $D/\mu=k_BT/e$, where $k_B$ is the Boltzmann constant. This method is adequate because the experimental data have been obtained in the limit of zero electric field [11]. The simulated system consists of 1 ion and 100 Ne atoms. The Newtonian equations of motions are numerically integrated with the Verlet algorithm [12] for suitable atom-atom and atom-ion interaction potentials. The time step for integration is set to $\tau=100\tau_i=2.42 \times 10^{-15}$ s, where $\tau_i=2\varepsilon_0 a_0 h/e^2$. $\varepsilon_0$, $a_0$, $h$, and $e$ are the vacuum permittivity, Bohr radius, Planck constant, and electron charge, respectively. After an initial equilibration time interval of $10^4$ time steps, the system evolves for $10^6$ more time steps for a total simulation time of 2.4 ns. Positions and velocities of all particles are recorded every 500 time steps, i.e., every 1.2ps, a typical time scale for the dynamics in dense systems [12]. Although the simulation should be carried out in the microcanonical ensemble (i.e., at constant energy), we did not observe any differences if the simulation is performed in the canonical ensemble (i.e., at constant $T$). For this reason, in order to keep the system temperature constant within a reasonable interval ($\pm$ 1 K), the particle velocities are rescaled every 20ps so as to keep the system average kinetic energy within the given limit. The validity of this MD method has been tested by reproducing the known diffusion coefficient of pure Ar [13], for which an accurate interatomic potential can be computed from the known multipole moments [14,15].

The situation is not as good for the ion-neutral interaction potential. In literature there are neither experimental determinations nor theoretical predictions for it. The long-range attractive tail of the potential can be computed from the knowledge of the dispersion coefficients $C_{2n}$ of the neutral and of the ion polarizability according to [14,15]. The repulsive part is completely unknown. There is only an indication that it could be modeled as the interaction of the neutral species with an infinite wall whose local charge density equals the electronic charge density of the ion [16]. In this case, the potential is proportional to the ion charge density [17] that is computed from the molecular eigenfunctions [18]. The charge density is fitted to an exponential form so as to obtain a Born-Mayer potential type $V_{rep}(r)=A\exp(-br)$. The potential strength $A$ is completely unknown and only a very crude estimate, within a factor 3, is given for He [16]. For the $O_2^-$-Ar system [9], we obtained an estimate for $A$ by a rough analogic procedure by inverting the dilute-gas mobility data of the systems $O_2^+$-He and $O_2^+$-Ar and $O_2^-$-He [19,20]. For the present $O_2^-$-Ne case, we again used an analogic procedure by looking at how the strength $A$ evolves for the $O^-$-rare gas neutral as a function of the neutral species [21] and interpolating between the previous estimates of $A$ for the $O_2^-$ species. As a result, the interaction potentials used in this MD study are shown in Fig. 2. In particular, the ion-neutral potential shows a hard-sphere radius of 0.35 nm, and its well is located at 0.40 nm and is 12 meV deep.

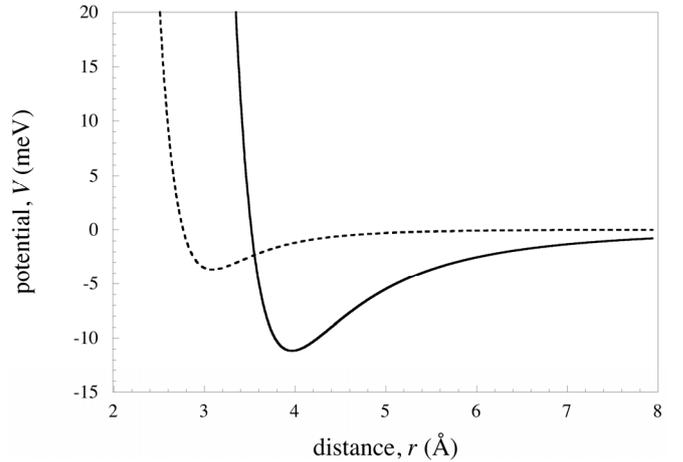

Fig. 2. $O_2^-$-Ne (solid line) and Ne-Ne (dashed line) interaction potentials.

## III. MD RESULTS

In each simulation only 1 ion is present, compared with the 100 neutral atoms. So, the statistics of the ionic properties is extremely poor. Thus, for each $N$ we have carried out 100 simulations starting with different initial positions and velocities of the particles picked from a randomly generated distribution complying with the requirement to be an equilibrium distribution at the desired $T$. Typical results for the mean square displacement $<r^2>$ of the ion are shown in Fig. 3 for $T$=45 K and $N$=10 nm$^3$. Owing to the poor statistics, the time evolution of $<r^2>$ is quite erratic in each single run (thin lines). The $<r^2>$ averaged over 100 runs is shown as a thick line. The diffusion coefficient $D$ is obtained by the linear fit $<r^2>=6Dt$. The diffusion coefficient is the converted to mobility by means of the Nernst-Townsend-Einstein relationship $\mu=eD/k_BT$. Finally, as it is customary, the density-normalized mobility $\mu N$ is computed. The MD simulation results are plotted and compared with the experimental data in Fig. 1.

The MD simulation results are well within a factor of 2 from the experimental data and reproduce their behavior relatively well. There is a region for $N<10$ nm$^{-3}$ in which $\mu N$ is quite constant and for $N>10$ nm$^{-3}$ $\mu N$ increases with $N$ as the experimental data do. The weak dip shown by the experimental data around $N$=11.5 nm$^{-3}$, probably related to the nearness of the critical density [4], is not reproduced by MD mainly because the correlation among particles, which is relevant near the critical point, is neglected in this approach. In any case, also by considering the uncertainty of the ion-atom interaction potential determination we can conclude that MD simulations yield mildly satisfactory results.

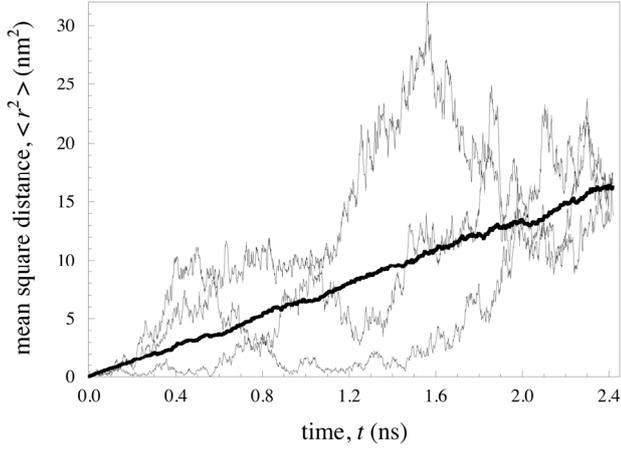

Fig. 3. Ion mean square displacement in single simulation runs (thin lines) and the average over 100 runs (thick line) for $T$=45 K and $N$=10 nm$^{-3}$.

In spite of this, MD lacks to clearly identify the physical reasons of the mobility behavior. These are to be sought by analyzing all pieces of information MD provides. First of all, we show in Fig. 4 the ion-atom pair correlation function $g(r)$ for some $N$. $g(r)$ gives the probability to find a neutral atom at distance $r$ from the ion. The low-density $g(r)$ shows only one peak at the position of potential well minimum and no minima at larger distance. By contrast, at higher densities, $g(r)$ develops a minimum for $r$= 0.54 nm followed by a secondary maximum at $r$=0.65 nm indicating the formation of the first solvation shell. The fact that $g(r)$ does not vanish at the solvation shell boundary suggests that there may be a dynamical interchange of atoms across it. A weaker second solvation shell is apparently building up, too.

The ion mainly interacts with its nearest neighbors, i.e., with the atoms in the first solvation shell. Their probability distribution at high density is well described by a Gaussian function, as shown in Fig. 5 for $N$=20 nm$^{-3}$. Their residence times follow an exponential distribution, as shown in Fig. 6 for the same density. These distributions are the characteristic signature of a memoryless Markovian process that is to be expected for pure Brownian motion.

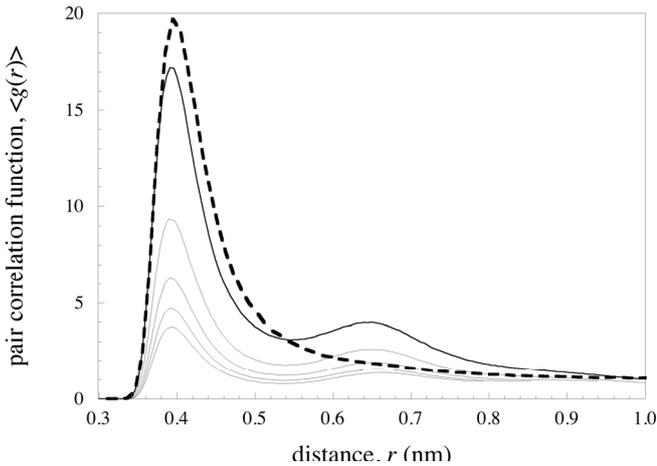

Fig. 4. Pair correlation function $g(r)$ for $N$=0.5 (thick dashed line), 5, 10, 15, 20, and 25 nm$^{-3}$ (thin lines, from top).

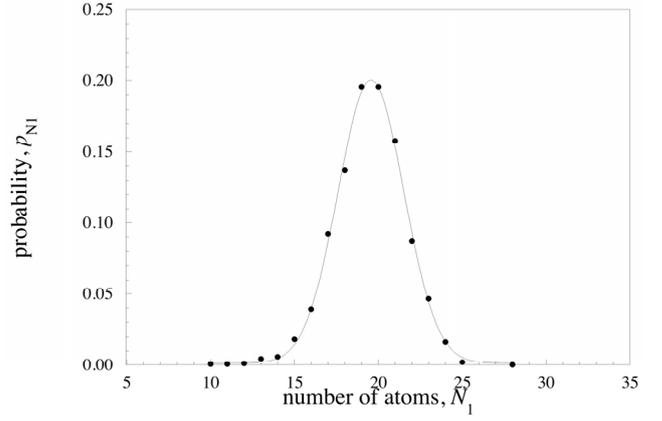

Fig. 5. Probability distribution of the number of Ne atoms within the first solvation shell for $T$=45 K and $N$=20 nm$^{-3}$.

It is reasonable to assume that the ion mobility is primarily determined by the average number $N_1$ of nearest atoms with which the ion most strongly interacts. $N_1$ is a weakly increasing function of the density $N$, as shown in Fig. 7, varying from approximately 18.5 to 20.5 in the range in which $N$ is increased by a factor of 5. Additionally, their average residence time, ~7 ps, is roughly density independent. Thus, we expect that the ion mobility gently decreases with increasing $N$. Actually, in the 5 to 10 nm$^{-3}$ range, in which the increase of $N_1$ with $N$ is steeper, $\mu N$ is roughly constant. For $N$>10 nm$^{-3}$ the mobility decrease due to the slower increase of $N_1$ is overcompensated by the increase of $N$ and the density normalized mobility further increases.

## IV. Conclusions

Equilibrium Molecular Dynamics simulations have proven useful to help reaserchers in describing and understanding the drift mobility of negative ions in a dense gaseous environment that is too dense for Kinetic Theory to apply and too dilute for a full hydrodynamic approach to be valid. MD simulations are straightforward to code and they provide scholars many properties of the ion-gas systems, such as the mean square displacement from which the diffusion coefficient is computed, the pair correlation function, and so on.

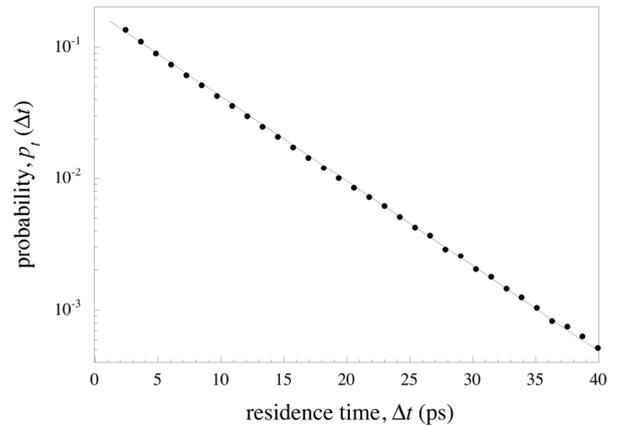

Fig. 6. Residence time probability distribution of Ne atoms within the first solvation shell for $T$=45 K and $N$=20 nm$^{-3}$.

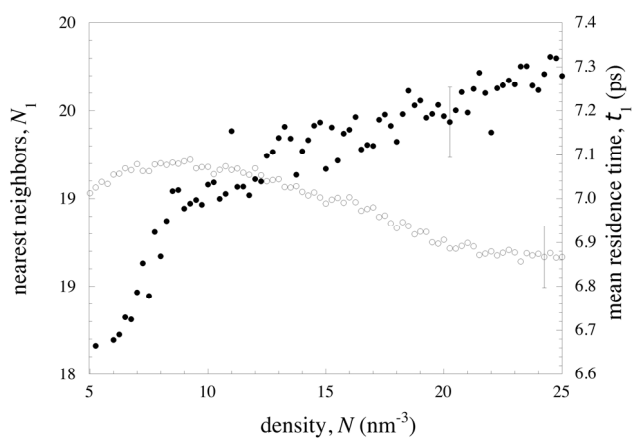

Fig. 7. Average number of nearest neighbors $N_1$ and average residence time $t_1$ of atoms in the first solvation shell. Error bars ars shown for only one point in order not to clutter the figure.

MD simulations also elucidate the possible formation of solvation shells around the ions if the density and temperature are favorable. We have successfully carried out MD simulations to explain the experimental behavior of the $O_2^-$ ion mobility in very dense Ne gas on an isotherm close to the critical one. The mobility data are reproduced quite well, in spite of the fact that the ion-atom interaction potential has only been determined by means of an educated guess owing to the complete absence of any literature suggestions. The MD results indicate that at high density there is the buildup of a first solvation shell and there is also an indication that a second solvation shell might be formed. The mobility appears to primarily determined by the momentum exchange with the nearest neighbor atoms in the first solvation shell around the ion with which the ion most strongly interacts.

We believe that the present results could be improved if a better knowledge of the ion-atom potential were available. Until some theoretical or experimental determination of the interaction potential, especially of its repulsive part, is produced, the only possible approach is to treat the strength of the Born-Mayer form of the repulsive potential as a free parameter and adjust its value until the best agreement with the mobility data is obtained. Unfortunately, this approach is difficult to pursue because of the heavy computational burden required by MD simulations of good enough statistical relevance.

REFERENCES

[1]  E. A. Mason, and E. W. McDaniel, Transport Properties of Ions in Gases, New York: Wiley, 1988
[2]  A. K. Bartels, "Density Dependence of Electron Drift Velocities in Helium and Hydrogen at 77.6 K", Appl. Phys., vol. 8, 1975, pp. 59-64.
[3]  J. A. Jahnke, M. Silver, and J. P. Hernandez, "Mobility of excess electrons and O2- formation in dense fluid helium", Phys. Rev. B, vol. 12, 1975, pp. 3420- 3427.
[4]  A. F. Borghesani, D. Neri, and M. Santini, "Low-temperature $O_2^-$ mobility in high-density neon gas", Phys. Rev. E, vol.48, 1993, pp. 1379-1389.
[5]  A. F. Borghesani, F. Chiminello, D. Neri, and M. Santini, "$O_2^-$ Ion Mobility in Compressed He and Ne gas", Int. J. Thermophys., vol. 16, 1995, pp. 1235–1244.
[6]  A. F. Borghesani, D. Neri, and A. Barbarotto, "Mobility of $O_2^-$ ions in near critical Ar gas", Chem. Phys. Lett., vol. 267, 1997, pp. 116–122.
[7]  K. F. Volykhin, A. G. Khrapak, and W. F. Schmidt, "Structure and mobility of negative ions in dense gases and nonpolar liquids", J.E.T.P., vol. 81, 1995, pp. 901-908.
[8]  H. G. Tarchouna, N. Bonifaci, F. Aitken, A. G. Mendoza Luna, and K. von Heften, "Formation of Positively Charged Liquid Helium Clusters in Supercritical Helium and their Solidification upon Compression", J. Phys. Chem. Lett., vol. 6, 2016, pp. 3036-3040.
[9]  A. F. Borghesani, "Mobility of $O_2^-$ ions in supercritical Ar gas: Experiment and molecular dynamics simulations", Int. J. Mass Spectrom., vol. 227, 2008, pp. 220-222.
[10] L. G. H. Huxley and R. W. Crompton, The Diffusion and Drift of Electrons in Gases, New York: Wiley, 1974.
[11] A. D. Koutselos and J. Samios, "Dynamics and drift motion of $O_2^-$ in supercritical argon", Mol. Liq., vol. 205, 2015, pp. 115-118.
[12] M. P. Allen and D. J. Tildesley, Computer Simulations of Liquids, Oxford: Clarendon Press, 1992.
[13] N. Gee, S.S.-S. Huang, T. Wada, and G. R. Freeman, "Comparison of transition from low to high density transport behavior for ions and neutral molecules in simple fluids", J. Chem. Phys., vol. 77, 1982, pp. 1411-1416.
[14] K. T. Tang and J. P. Toennies, "The van der Waals potentials between all the rare gas atoms from He to Rn", J. Chem. Phys. 2003, pp. 4976-4983.
[15] G. C. Maitland, M. Rigby, E. B. Smith, and W. A. Wakeham, Intermolecular Forces. Their origin and determination, Oxford: Clarendon Press, 1981.
[16] J. A. Barker and I. P. Batra, "Corrugation studies for p (2x2) and c (2x2) phases of oxygen on Ni(001) and Cu(001)", Phys. Rev. B, vol. 27, 1983, pp. 3138-3143.
[17] N. Esbjerg and J. K. Norskov, "Dependence of the He-Scattering Potential at Surfaces on the Surface-Electron-Density Profile", Phys. Rev. Lett., vol. 45, 1980, pp. 807-810.
[18] P. E. Cade and A. C. Wahl, "Hartree-Fock-Roothan wavefunctions for diatomic molecules. First-row homonuclear systems $A_2$, $A_2^{\pm}$, and $A_2^{*}$", At. Data and Nucl. Data Tables, vol. 13, 1974, pp. 339-389.
[19] I. Dotan, W. Lindinger, and D. L. Albritton, "Mobilities of various mass-identified positive and negative ions in helium and argon", J. Chem. Phys., vol. 64, 1976, pp.4544-4547.
[20] W. Lindinger and D. L. Albritton, " Mobilities of various mass-identified positive ions in helium and argon", J. Chem. Phys., vol. 62, 1975, pp.3517-3522.
[21] L. A. Viehland and C. C. Kirkpatrick, "Relating ion/neutral reaction rate coefficients and cross-sections by accessing a data base for ion transport properties", Int. J. Mass Spectrom. Ion Proc., vol. 149/150, 1995, pp. 555-571.